\documentclass[12pt]{article}

\usepackage{epsfig}
\usepackage{amsfonts}

\newcommand{\beq}{\begin{equation}}    
\newcommand{\eeq}{\end{equation}}

\catcode`\@=11
\newcommand{\bs}[7]{\bibitem{#1#2} #3, 19#2. #4. #5,  #6:#7.}

\newcommand{\bbo}[5]{\bibitem{#1#2} #3, 19#2. #4. #5.}
\def\ZZ{Z\kern -5pt Z}
\def\RR{R\kern -5pt R}
\def\NN{N\kern -5pt N}

\begin{document}

\vskip 1.0 cm

\title{\sc Finite time and asymptotic behaviour \\ 
of the maximal excursion of a random walk}
\vskip 0.5cm
\author{\sc\small Roger Bidaux\thanks{bidaux@spec.saclay.cea.fr}, 
J\'er\^ome Chave\thanks{chave@spec.saclay.cea.fr, to whom 
correspondence should be addressed}\ \   
and Radim Vo\v{c}ka\thanks{radim@spec.saclay.cea.fr} \\ \\
{\em\small Service de Physique de l'Etat Condens\'e} \\
{\em\small SPEC/DRECAM, CEN Saclay} \\
{\em\small F-91191 Gif-sur-Yvette, France}}
\date{}
\maketitle

\begin{abstract}
We evaluate the limit distribution of the maximal excursion of a random walk
in any dimension for homogeneous environments and for self-similar supports 
under the assumption of spherical symmetry. This distribution is obtained in 
closed form and is an approximation of the exact distribution comparable to 
that obtained by real space renormalization methods.
Then we focus on the early time behaviour of this quantity.
The instantaneous diffusion exponent $\nu_n$ exhibits a systematic overshooting
of the long time exponent. Exact results are obtained in one dimension up to 
third order in $n^{-1/2}$. In two dimensions, on a regular lattice and on the 
Sierpi\'nski gasket we find numerically that the analytic scaling 
$\nu_n \simeq \nu+A n^{-\nu}$ holds. 
\\
{\bf Keywords}: random walk, maximal excursion, finite size scaling, 
enumeration technique, Sierpi\'nski gasket
\end{abstract}

\vskip 1cm

The random walk (RW) on a lattice has long been
studied due to its widespread applications in mathematics, physics, 
chemistry and other research areas.
It turns out that despite the huge amount of
accomplished work, it still remains a thriving research topic.
Lots of results can be obtained in the continuum limit
(Brownian motion) but results for RW on a lattice often yield drastically 
different behaviour - as it is the case for the winding angle distribution 
\cite{Rudnick87} - or, at least, unusual finite time convergence properties.
In the present work we investigate a central quantity for RW, the maximal 
excursion from the origin at time $n$,
$M_n = \max(||x_m||, 0\leq m\leq n)$.
This random variable is of great interest in many practical purposes 
such as the control of pollution spread, propagation ranges of epidemics,
tracer displacement in fluids, the radius of gyration of polymer chains  
\cite{Rubin78, Whittington87},
or of lattice animals \cite{Privman84, Conway95} or other extreme statistics.
A great deal of work was devoted to the first passage time (FPT) statistics
which is a closely related quantity. Nevertheless methods used 
to find the exact FPT distribution in one dimensional 
inhomogeneous environments 
\cite{Murthy89, Bouchaud90, Raykin93, Chave99} do not help to get a 
closed form of the
exact distribution of $M_n$. Except for the one dimensional case,
only the leading order asymptotic expressions  
(as $n \to \infty$) are available. 
It was proved long ago by Erd\H{o}s and Kac \cite{Erdos46}, that in 
this limit the distribution of $M_n$ coincides 
with that of the Brownian motion.
 This result appears as some kind of a central limit theorem. 
However, it does not deal with centered, reduced variables. 
Moreover it offers no practical access (for physically motivated 
purposes) to the convergence speed towards the limit law. 
The only global estimates available for $M_n$ are the laws of iterated 
logarithm of Khinchine and Chung \cite{Revesz90} for the one dimensional
random walk,
claiming that although all the distributions have the same limit form,
the intrinsic uncertainty on $M_n$ {\em increases} with $n$.
Hence, it is not clear what the  finite time behaviour of the maximal 
excursion $M_n$ is. It is our aim to clarify this point.

In this article, we first derive the leading order expression for the 
distribution of $M_n$ in a generalized form. This expression is shown to
apply also on self-similar structures.
Then we proceed to the next leading order expansion for short time. In this
regime, the first moment of $M_n$ scales as 
$\left<M_n\right>\sim n^{\nu_n}$
where $\nu_n$ is the  effective instantaneous diffusion exponent, and
tends to $\nu_n \to \nu$ as $n \to \infty$ ($\nu = 1/2$ on regular lattices,
$\nu=\ln 2/\ln (d+3)$ on the $d$-dimensional Sierpi\'nski gasket). 
We show numerical evidence that the effective instantaneous diffusion
exponent $\nu_n$ approaches the limiting value $\nu$ according
to $\nu_n - \nu \sim n^{-\nu}$.
This result is valid for both regular and self-similar lattices.
We finally discuss this point in the context of other problems
of statistical physics. 

First let us briefly recall the formulae for the maximal excursion 
of a $d$-dimensional Brownian motion ${\bf r}_t$, that is  
$M_t = \max(||{\bf r}_u||_2, u\leq t)$, where 
$||{\bf r}||_2 = \sqrt{\sum_i r_i^2}$ is the Euclidean distance. 
The limit distribution is denoted by ${\bf P}_d(a,t) = {\rm Pr}\{M_t<a\}$. 
The calculation goes through the 
solution of the diffusion equation in $d$ dimensions with spherical symmetry
and absorbing boundaries on the hypersphere of radius $a$. 
Let $U({\bf r},t)$ be the probability density function for
the position vector ${\bf r}$ of the walker relative to the origin at 
time $t$, without ever crossing the hypersphere boundary at distance $a$.
Then $U({\bf r},t)$ satisfies the diffusion equation
\beq
\label{eq1}
\partial_tU({\bf r},t)=\frac{1}{2d}\nabla^{2}_{\bf r}U({\bf r},t)
\eeq
where $\nabla^{2}_{\bf r}$ is the $d$-dimensional Laplace operator. 
The diffusion constant is set as $1/(2d)$ so that the solution corresponds
to a simple random walk on ${\bf Z}^d$ with a time $\tau$ between steps
and a lattice spacing $\sqrt{\tau}$ in the limit $\tau \to 0$. 
The boundary condition is that $U({\bf r},t)=0$ for $||{\bf r}||_2=a$, and
the initial condition is
\[
U({\bf r},0)=\delta({\bf r})={\delta_+(|{\bf r}|) 
\over A_d|{\bf r}|^{d-1}}\,,\]
where $A_d$ is the surface area of the unit hypersphere in $d$ dimensions
and $\delta_+$ is the (one-sided) delta function. The probability of
remaining inside the hypersphere up to time $t$, ${\bf P}_d(a,t)$, 
is the volume integral
of $U({\bf r},t)$ over the hypersphere. Due to spherical symmetry, 
$U({\bf r},t)$
is a function of $r=||{\bf r}||_2$ only, which we now denote by $U(r,t)$,
so that, from (\ref{eq1})
\beq
\partial_t U(r,t) = {1\over 2d r^{d-1}}
\partial_r r^{d-1}\partial_r U(r,t),
\label{eqdif}
\eeq 
with  $U(r,0)={\delta_{+}(r)\over A_d r^{d-1}},~U(a,t)=0$ and
\[
{\bf P}_d(a,t) = \int_0^a A_d r^{d-1} U(r,t)dr
\]
The solution of (\ref{eqdif}) is given in the form of an infinite 
eigenfunction expansion.
This calculation can be done for self-similar lattices in the framework
of the O'Shaughnessy-Procaccia approximation \cite{OSaughnessy85}.
It consists in assuming a spherical symmetry of a fractal object,
and in introducing an effective diffusion coefficient $D=D_0 r^{-2+1/\nu}$
computed from the solution of the stationary diffusion problem on 
self-similar lattices without angular dependence. 
Thanks to this approximation, an analytic approach can be pursued.
The final distribution, denoted ${\bf P}_{d,\nu}$
for general $\nu$ is obtained in the Laplace domain in closed form
\beq
{{\bf P}_{d,\nu}}(a,s) = {1\over s}\left[1-{2^{1-d\nu}
\over \Gamma(d\nu)}
{(4\nu^2D_0^{-1}a^{1\over\nu}s)^{d\nu-1\over 2}\over 
I_{d\nu-1}\left(\sqrt{4\nu^2D_0^{-1}a^{1\over\nu}s}\right)}\right]
\label{LD}
\eeq
Here $I_n(x)$ is the modified Bessel function of order $n$.
From this formula all moments are plainly computed
\beq
\langle M^k_t\rangle_{d,\nu}=\left\{{2\nu k\over\Gamma(k\nu+1)}
{ 2^{1-d\nu}\over \Gamma(d\nu)(4D_0^{-1}\nu^2)^{k\nu}}
\int\limits_{0}^{\infty}\frac{u^{(2k+d)\nu-2}}{I_{d\nu-1}(u)}du\right\}
t^{k\nu}
\label{momsier}
\eeq
Putting $D_0=1/2d$ and $\nu=1/2$ in (\ref{LD}), we easily recover the known
distributions on regular lattices in one \cite{Erdos46}, two \cite{Seshadri80}
and three \cite{Rubin78} dimensions. A similar method 
was used in \cite{Koplik94} to solve the first passage time problem in
the presence of a steady potential flow.
It is worthwhile mentioning that the limit $d\rightarrow \infty$ 
for $\nu = 1/2$ in (\ref{LD})  yields 
${{\bf P}_\infty}(a,s) = \frac{1}{s}\left(1-\exp(-a^2s)\right)$
to leading order.
Hence, as intuitively expected, the maximal excursion of a random walker 
is exactly known in infinite dimension and peaks at $a=\sqrt{t}$. 

\begin{figure}[h]
\begin{center}
\epsfig{file=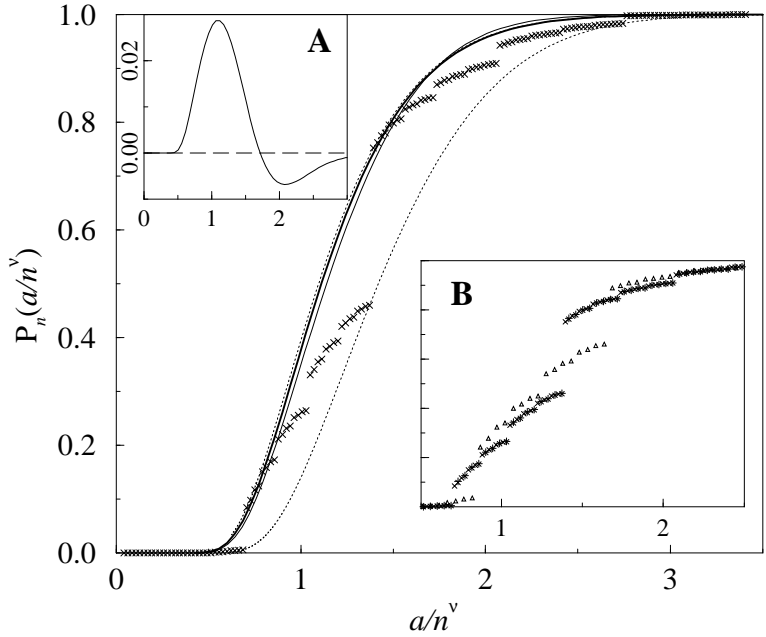,width=10cm}
\caption{Rescaled distribution of the maximum excursion 
$P_n(a/n^{\nu})$ versus the reduced variable $r/n^{\nu}$ for $n=7500$ 
($\times$) on the two dimensional Sierpi\'nski gasket, compared to
the analytical prediction (\ref{LD}), bold line, and to the RSRG result, 
thin line. Dotted lines: $P_n(a/n^{\nu})$ for fixed $a$ and
varying $n$: the rightmost curve is computed for $a=2^5$ while 
the leftmost curve corresponds to $a=2^5+1$. Inset A: difference between 
(\ref{LD}) and the RSRG prediction. Inset B: $P_n(a/n^{\nu})$ for $n=1000$
($\triangle$) and $n=1500$ (+). The curve for  $n=7500 = 5\times 1500$ 
($\times$) is exactly superposed to the curve for $n=1500$.}
\label{dist}
\end{center}
\end{figure}

On self-similar lattices ($\nu\ne 1/2$), Equation (\ref{LD}) is only an
estimate of the exact limit law. 
We have compared distribution (\ref{LD}) with
the distribution obtained by real space renormalization group (RSRG) 
techniques for the 2D Sierpi\'nski gasket \cite{Broeck89,Yuste95, Porra98}.
Both laws turn out to approximate the exact
distribution to the same order (see Figure \ref{dist} and the discussion
in \cite{Klafter91}). 
We have also evaluated the moments. To first order, they behave 
as $\langle M^k\rangle \sim n^{k\nu}$, so we define the {\em normalized}
moments $\langle M^k\rangle = \langle M^k_n\rangle/n^{k\nu}$ which tend
to a constant asymptotically.
The first two normalized moments obtained from (\ref{momsier}), 
$\langle M\rangle=1.20$ and $\langle M^2\rangle=1.59$, 
should be compared with the moments obtained
from the RSRG method ($\langle M\rangle=1.19$, 
$\langle M^2\rangle=1.57$) and with the numerical estimates 
($\langle M\rangle=1.28$, $\langle M^2\rangle=1.84$). 
Both theoretical formulae underestimate the actual values \cite{Klafter91}.
This can be understood as follows.
Strictly speaking, no limit distribution can be defined for $M_n$
but, for consecutive time series $n, 5n, 5^2n,\ldots$, $P_n(a/n^{\nu})$ 
is left invariant, because if a random walker takes $T$ steps
to leave a triangle of size $R$, it needs a time $5T$ to leave a  
triangle twice bigger. 
Thus $P_n(a)$ fulfills the scaling relation $P_n(a)=P_{5n}(2a)$. 
However, between these times and for fixed $a/n^{\nu}$, the rescaled 
distribution $P_n(a/n^{\nu})$ has a log-periodic variation.
This log-periodic behavior has been known for some time 
for lattice random walks \cite{hugues95}.
Both function  (\ref{LD}) and the RSRG
result give the probability to stay in the triangle of size $R=2^i$, 
that is one minus the probability to reach sites at distance $R+1$ from
the origin. As observed in Figure (\ref{dist}) this 
corresponds to an extremum in the oscillation of $P_n(a/n^{\nu})$ 
(leftmost dashed line of Fig. \ref{dist}) rather than to an average.

Now we turn our attention to the convergence speed towards the
asymptotic law (\ref{LD}).
For convenience we study the case of the discrete
time random walk on the lattice, where analytical results
can be obtained in one dimension and an exact numerical approach
is possible in higher dimensions. 
We focus on the instantaneous diffusion exponent $\nu_n$
which furnishes an information about the convergence speed of the 
moments. The numerical estimation of $\nu$ using a Monte-Carlo sampling 
can lead to false conclusions as in \cite{dekeyser87} (see \cite{manna89}). 
Here we use exact enumeration methods and therefore 
we avoid such problems.

The exact solution of the problem in one dimension 
is obtained by solving the master equation 
with absorbing boundaries at points $\pm a$ with the use 
of a Fourier development (obtained in \cite{Weiss94} with a minor misprint),
but the moments cannot be calculated in a straightforward manner from this
expression beyond first order.
We derived another form of the distribution by a recursive use of the 
reflection theorem. The probability density of the maximal 
excursion at step $n$ reads 
\beq 
\label{fdens}
P_n(M_n=a)=2\sum_{k=0}^{\infty}(-1)^k\biggl[p_n\Bigl((2k+1)a\Bigr) + 
p_n\Bigl((2k+1)(a+1)\Bigr)+2\sum_{i=1}^{2k}p_n\Bigl((2k+1)a+i\Bigr)\biggr] 
\eeq 
where $p_n(x)$ is the probability density function for the discrete 
random walk to be at position $x$ (which is null for $|x|>n$).
Formula (\ref{fdens}) allows a convergent expansion of the first moments
in powers of $n^{-1/2}$, which exist, since $P_n(M_n=a)$ is an analytic 
function of $n^{1/2}$.
At order $n^{-1/2}$, divergent series are encountered which can be 
summed by classical methods \cite{Hardy49}, yielding
\beq 
\langle M_n\rangle =  
\sqrt{\pi n\over 2}-{1\over 2}+{1\over 12}\sqrt{2\pi\over n}+
{\mathcal O}\left({1\over n}\right)
\label{theorie}
\eeq 
\beq 
\langle M_n^2\rangle = 2G n-\sqrt{\pi n\over 2}+{G+1\over 3}+
{\mathcal O}\left({1\over \sqrt{n}}\right)
\label{theorie2m}
\eeq
where $G = 0.9166\ldots$ is the Catalan constant.
In the calculation of the second cumulant the terms of order $\sqrt{n}$ and 
$1/\sqrt{n}$ cancel, as expected. The distribution $P_t(M_t=a)$ 
for continuous time random walks (CTRW) follows plainly from (\ref{fdens})  
since $P_t(M_t=a)=\sum_n P_n(M_n=a)\Pi_t(n)$, where $\Pi_t(n)$ is the 
probability for the CTRW to perform $n$ steps in time $t$. 
Numerically, a series expansion similar to  (\ref{theorie2m}) is found 
for an exponential distribution of waiting times. The exponential 
distribution is particular because it
is the only one for which a master equation formulation and 
a CTRW on the same lattice are isomorphic \cite{kenkre73}.
A striking feature of the random variable $M$ compared to other extremal
quantities at finite times is that the leading order expansion 
of $\langle M_n^k\rangle$ scales as  $n^{k/2} + {\rm cte.} n^{(k-1)/2}$
and not as $n^{k/2} + {\rm cte.} n^{k/2-1}$, hence finite size effects 
persist for a large number of steps. 

\begin{figure}[h]
\begin{center}
\epsfig{file=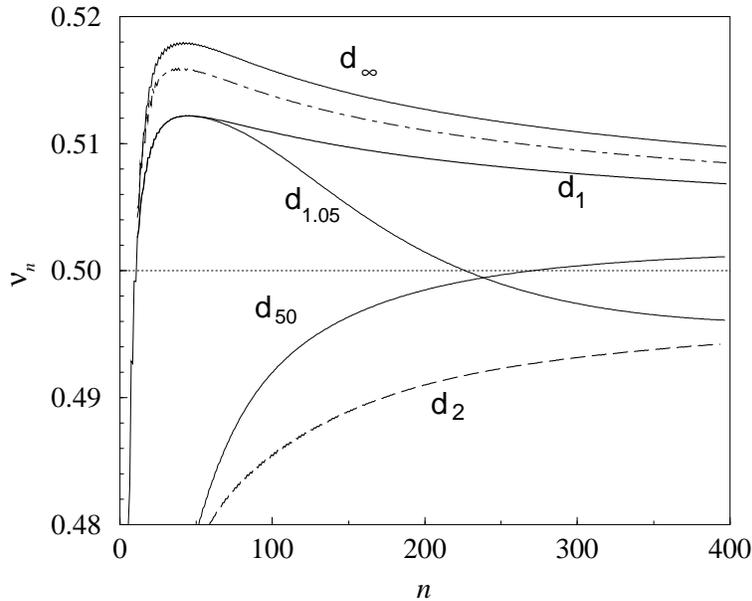,width=10cm}
\caption{Figure 2: Instantaneous exponent $\nu_n$ 
(averaged over two consecutive steps) 
versus number of steps on the square lattice at two dimensions.
Enumeration up to step $n=400$ in metric $d_1$ (bold solid line)
and metric $d_\infty$ (solid line). The one dimensional situation (caliper
diameter) is given for reference (dot-dashed line). $\nu_n$ is also computed
using the metric $d_2$ (Euclidean distance, dashed line), 
$d_{1.05}$ and $d_{50}$ (solid lines).}
\label{nu2d}
\end{center}
\end{figure}

\begin{figure}[h]
\begin{center}
\epsfig{file=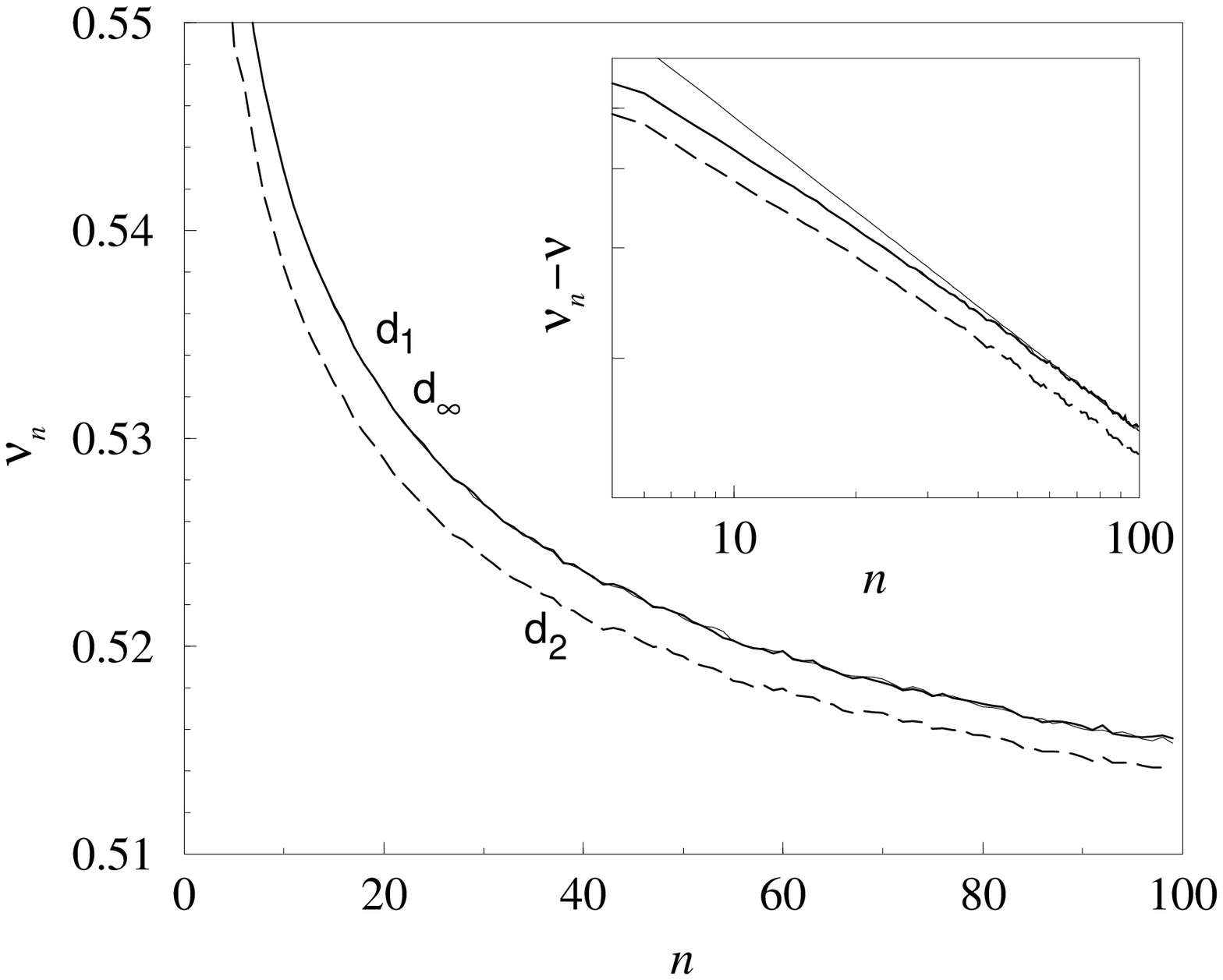,width=10cm}
\caption{Instantaneous exponent $\nu_n$
versus number of steps in the off-lattice model at two dimensions.
Monte-Carlo simulation up to step $n=100$ in metric $d_1$ (bold solid line)
$d_2$ (dashed line), and $d_\infty$ (solid line). The results for metric
$d_1$ and $d_\infty$ are almost indistinguishable, as expected. 
Inset: Log-log plot of 
$\overline \nu_n - \nu$ which shows the  $n^{-1/2}$ scaling.}
\label{nuMC}
\end{center}
\end{figure}

In two dimensions, no exact result is available for finite times
and the analyticity of the probability density $P_n(M_n=r)$ is not obvious.
Hence we investigate this case numerically. It is possible to perform an
exact enumeration of the walks by studying the joint probability density
of the position and maximal excursion $P_n(x,y,M)$ on the square lattice 
${\mathbb Z}^2$. We can compute $P_n(x,y,M)$ in the region 
$0\leq x\leq M,~0\leq y\leq x$ only, due to symmetries. 
We use the family of metrics 
$d_p({\bf x}) = \left(\sum_{i}(|x_i|^p)\right)^{1/p}$
to compute the maximal excursion from the origin of the lattice.
In Figure \ref{nu2d} we plot the instantaneous exponent $\nu_n$
with three classical choices of metric:  $d_1$, $d_2$ (Euclidean distance) 
and $d_\infty$ (max distance).
The metric $d_1$ and $d_\infty$ both induce a strong overshooting
of $\nu_n$ with the limit value $1/2$. The curves have the same shape 
as that of the one dimensional case, also potted for reference in
Figure \ref{nu2d}. 
The first two moments have many features in common with their 
1D equivalents. For example, using the metric $d_\infty$ we find that 
the series expansion 
$\langle  M_n^k \rangle = \sum_{p=0}^{\infty}m^k_{k-p} (\sqrt{n})^{k-p}$
holds for both first and second moments up to third order, 
with $m^1_0 = -0.50,~m^1_{-1} = 0.322$ and 
$m^2_{1} = -1.083,~m^2_0 = 0.95,~m^2_{-1}=-0.33$.
The leading order terms are exactly evaluated from the 
asymptotic results and read $m^1_1 = 1.0830,~m^2_2 = 1.3048$.
In the Euclidean metric $d_2$, we find a drastic change in the shape
of the curve $\nu_n$ (Figure \ref{nu2d}). 
The instantaneous exponent approaches ${1\over 2}$ 
from {\em below} and remains below ${1\over 2}$ at time $n=400$. 
This phenomenon is a lattice effect. We show this fact
by computing $\nu_n$ in an off-lattice random walk model (Figure \ref{nuMC}).
Since it is not possible to use exact enumeration techniques in this case,
we resort to a Monte-Carlo simulation. We inspect $2.10^8$
random walks with fixed distance increments and a uniform distribution 
of the angles (Pearson walks \cite{Weiss94}). In this situation, $\nu_n$ 
in metric $d_2$ is very close to that obtained in metric 
$d_1$ and $d_\infty$, and it {\em decreases} towards ${1\over 2}$.
 Both lattice and off-lattice models
should give equivalent results once the discretization effects are 
smoothed out. Therefore $\nu_n$ should ultimately approach $1\over 2$ 
from above in the on-lattice model.
We have investigated the change of $\nu_n$ when varying 
continuously the metric $d_p$ with $1\leq p \leq \infty$ on the 
lattice (Figure \ref{nu2d}). For large enough values of $p$ ($p>50$),
we do observe that the curve crosses the value ${1\over 2}$. 
In the metric $d_2$,
however, the time needed for $\nu_n$ to cross ${1\over 2}$ should be enormous.

This result shows that the definition of the metric strongly influences
the convergence properties of the maximal excursion on regular lattices. 
The two natural metrics for the square lattice, $d_1$ and $d_\infty$,
lead to a  behaviour similar to that observed in the continuum model. 

\begin{figure}[h]
\begin{center}
\epsfig{file=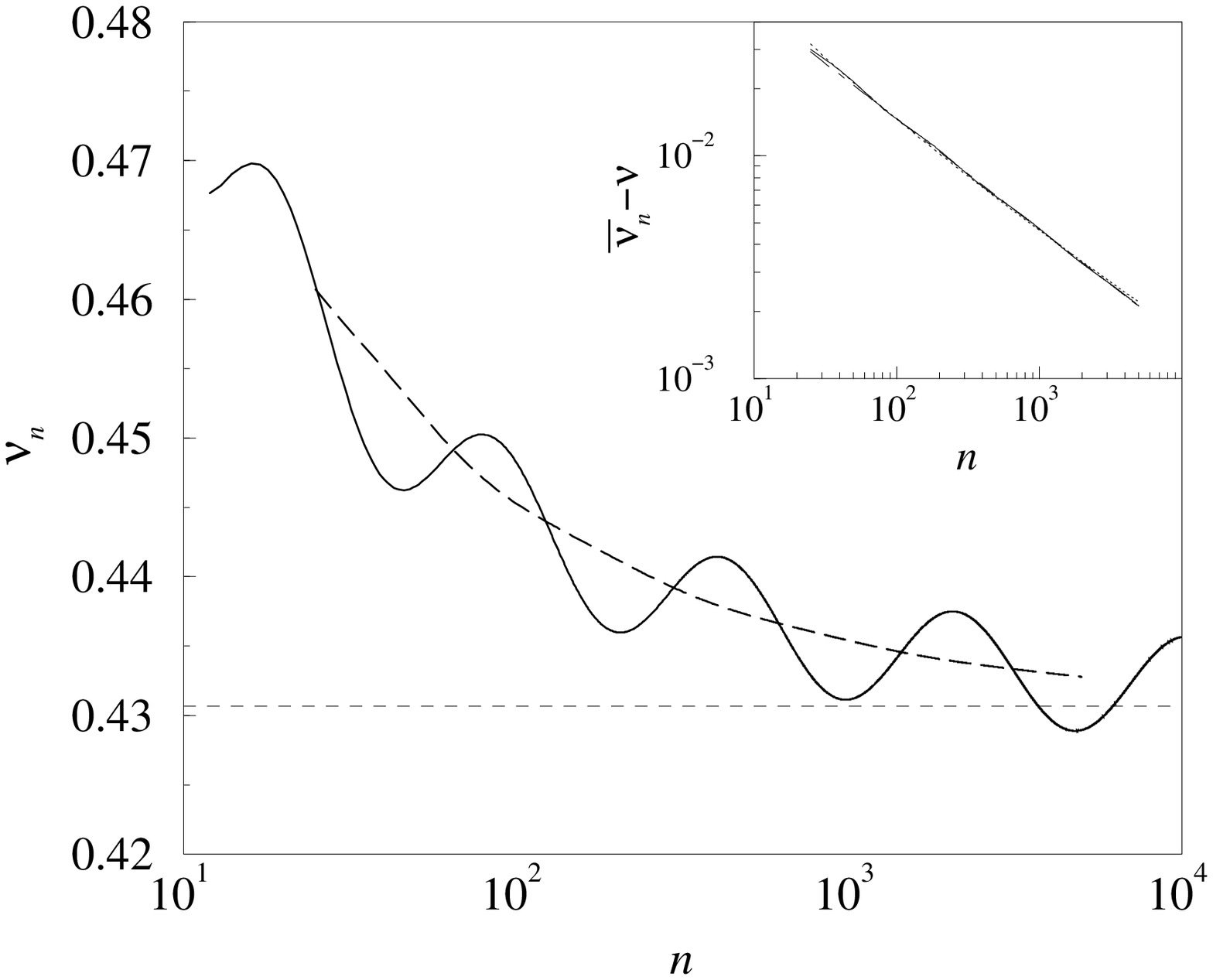,width=10cm}
\caption{Convergence of the instantaneous exponent 
for the first moment $\nu_n$ (bold line) towards the limit value 
$\ln 2/\ln 5$ (dashed line) on the two dimensional Sierpi\'nski gasket. 
The running average $\overline \nu_n$ is also plotted (dashed bold line).
Inset: Log-log plot of $\overline \nu_n - \nu$ and fit (\ref{fitt}).}
\label{nusierp}
\end{center}
\end{figure}

On the Sierpi\'nski gasket we enumerate the walks starting from the top 
of the biggest triangle up to a {\em fixed} time. The metric chosen here is
the chemical distance from the origin. For each 
maximal excursion $r$ we compute the probability of remaining below $r$
after $n$ steps, $P_n^S(r)$. Unlike the transfer matrix method, this method 
works only at fixed time, but allows us to discard the long
transient regime and to compare the exact limit distribution to 
its spherically symmetric approximation. 
We have computed the instantaneous exponent $\nu_n$ up to step $n=10^4$ on a 
gasket of size 256 (cf. Figure \ref{nusierp}).
Like the moments, $\nu_n$ displays a log-periodic 
oscillation persisting in the long time regime  with an amplitude 
less than $8~10^{-3}$.
$\nu_n$ tends to the asymptotic value $\nu = {\ln 2\over \ln 5}$ for long time.
It seems that on a very general class of lattices
the finite time behaviour of $\nu_n$ and therefore of $\langle M^k_n \rangle$ 
is an analytic function of $n^{-\nu}$. 
This fact lacks a clear physical understanding. The real space renormalization 
results do show that $n^{\nu}$ is the well defined time scale for this 
problem but the exact evaluation of finite size effects is not accessible
from this method. To assess this hypothesis we have smoothed out the 
log-periodic oscillations of $\nu_n$.
For a log-periodic function $f(x) = f(Tx)$, one can define $z = \ln(x)$
and $\tilde f(z) = f(x)$ so that the running logarithmic average reads
\[
\overline f\left({Tx\over 2}\right) = {1\over \int_z^{z+\ln T} du} 
\int_z^{z+\ln T} \tilde f(u) du
= {1\over \int_x^{Tx} {dv\over v}} \int_x^{Tx} f(v) {dv\over v}
\]
Using a discrete form of this average we write 
\[
\overline \nu_{{5n\over 2}} = 
{1\over \sum_{i=n}^{5n}{1\over i}}\sum_{i=n}^{5n}{\nu_i\over i} 
\]
We tried to fit $\overline \nu_n$ using
\beq
\overline \nu_n = \nu+A n^{-\nu}+B n^{-2\nu} + o(n^{-2\nu})
\label{fitt}
\eeq 
and we found a very good regression for $A=0.082\pm 0.002$ and 
$B=0.18\pm 0.1$. However, the best fit with only one power law 
is $\nu+{\rm cte}. n^{-\alpha}$ with $\alpha=0.49$, so that strictly speaking 
a nonanalytic short time dependence cannot be ruled out. 
The underlying assumption in the computation of $\overline \nu_n$ is that 
the regular log-oscillatory pattern of $\nu_n$ is additive.
This assumption does not hold because
the averaged exponent $\overline \nu_n$ still shows
some oscillation. The local exponent $\alpha$ fluctuates between 
0.40 and 0.52,
which does not allow to confirm unambiguously the hypothesis of the analytic 
behavior of $\overline \nu_n$ as a function of $n^{\nu}$. 

We would like to point out that in the context of 
lattice animals, the quantity $\langle M^k_n \rangle$, or equivalently the 
``caliper diameter'' (average spanning diameter of lattice animals once
projected on a fixed axis)
displays a sub-leading order behaviour which scales as $n^{(k-1)\nu}$
\cite{Privman84}, aside from the well-known non analytic subleading 
term, and can be interpreted as a 'surface contribution'. 
In the case of the maximal excursion of a random walk, we have 
proved in one dimension and evidenced through enumerations in higher
dimensions that the early time instantaneous exponent $\nu_n$ is
systematically above its limit value $\nu$ with
a leading order development 
$\nu_n\simeq \nu+A n^{-\nu}$ where $A$ depends on the 
precise choice of the metric. 
This result is consistent with the fact that the corrective 
scaling to the moments due to finite size effects includes only terms 
of the form $(n^{-\nu})^p,~p\in{\mathbb N}$, which was proved in 
one dimension and which can also be interpreted as a surface contribution. 

In conclusion, besides its intrinsic interest, the maximal excursion of a 
random walk shares difficulties which are often encountered in physical 
problems dealing with finite size series. 
In certain cases, power law exponents inferred from the finite size series 
expansions should be considered with caution, as might be the case for 
directed percolation series.

\newpage

\end{document}